\documentclass[12pt,english]{revtex4}
\usepackage[T1]{fontenc}
\usepackage[latin9]{inputenc}
\usepackage{float}
\usepackage{amsmath}
\usepackage{amssymb}
\usepackage{graphicx}
\usepackage{wasysym}

\makeatletter
\@ifundefined{textcolor}{}
{%
 \definecolor{BLACK}{gray}{0}
 \definecolor{WHITE}{gray}{1}
 \definecolor{RED}{rgb}{1,0,0}
 \definecolor{GREEN}{rgb}{0,1,0}
 \definecolor{BLUE}{rgb}{0,0,1}
 \definecolor{CYAN}{cmyk}{1,0,0,0}
 \definecolor{MAGENTA}{cmyk}{0,1,0,0}
 \definecolor{YELLOW}{cmyk}{0,0,1,0}
}

\makeatother

\usepackage{babel}
\begin{document}

\title{Phase response function for oscillators with strong forcing or coupling}

\author{{\normalsize{}Vladimir~Klinshov$^{1}$, Serhiy~Yanchuk$^{2}$,
Artur Stephan$^{3}$, and Vladimir~Nekorkin$^{1}$}}

\affiliation{$^{1}$Institute of Applied Physics of the Russian Academy of Sciences,
46 Ul'yanov Street, 603950, Nizhny Novgorod, Russia}

\affiliation{$^{2}$Technical University of Berlin, Institute of Mathematics,
Straße des 17. Juni 136, 10623 Berlin, Germany}

\affiliation{$^{3}$Humboldt University of Berlin, Unter den Linden 6, 1099 Berlin,
Germany}

\begin{abstract}
Phase response curve (PRC) is an extremely useful tool for studying
the response of oscillatory systems, e.g. neurons, to sparse or weak
stimulation. Here we develop a framework for studying the response
to a series of pulses which are frequent or/and strong so that the
standard PRC fails. We show that in this case, the phase shift caused
by each pulse depends on the history of several previous pulses. We
call the corresponding function which measures this shift the\emph{
phase response function} (PRF). As a result of the introduction of
the PRF, a variety of oscillatory systems with pulse interaction can
be reduced to phase systems. The main assumption of the classical
PRC model, i.e. that the effect of the stimulus vanishes before the
next one arrives, is no longer a restriction in our approach. However,
as a result of the phase reduction, the system acquires memory, which
is not just a technical nuisance but an intrinsic property relevant
to strong stimulation. We illustrate the PRF approach by its application
to various systems, such as Morris-Lecar, Hodgkin-Huxley neuron models,
and others. We show that the PRF allows predicting the dynamics of
forced and coupled oscillators even when the PRC fails. Thus, the
PRF provides an effective tool that may be used for simulation of
neural, chemical, optic oscillators, etc.
\end{abstract}
\maketitle
A variety of physical, chemical, biological, and other systems exhibit
periodic behaviors. The state of such a system can be naturally determined
by its phase \citep{Winfree2001}, that is, the single variable indicating
the position of the system within its cycle. The concept of the phase
proved to be exceptionally useful for the study of driven and coupled
oscillators \citep{Winfree2001,Kuramoto2012,Pikovsky2003}.

\begin{figure}
\begin{centering}
\includegraphics[bb=0bp 0bp 230bp 80bp,clip]{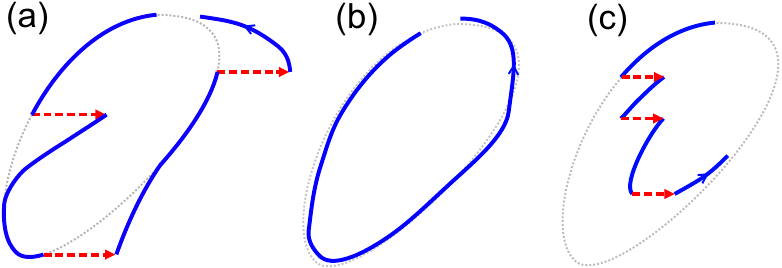}
\par\end{centering}
\caption{\label{fig:sketch}A forced oscillator: (a) strong but sparse forcing,
(b) continuous but weak forcing, (c) strong and frequent forcing.
Gray dotted curve is the stable limit cycle, blue lines denote trajectories
of the system, red dashed lines depict the action of the external
pulses.}
\end{figure}

In order to describe the response of oscillators to an external force
or coupling the so-called phase response curve (PRC) is widely used.
The PRC defines the oscillator's response to a single short stimulus
(pulse). The PRC can be calculated numerically or measured experimentally
for oscillatory systems of different origin \citep{Schultheiss2012}.
These properties made it a useful tool for the study of forced or
coupled oscillators \citep{Guckenheimer1975,Ermentrout1996,Brown2004,Chandrasekaran2011,Luecken2013,Luecken2012,Canavier2010},
and it is especially effective in neuroscience where the interactions
are mediated by pulses. If the pulse arrivals are separated by sufficiently
long time intervals, the transient caused by a pulse vanishes before
the next one comes. From the theoretical point of view, it means that
the system returns to the vicinity of its stable limit cycle before
the next pulse arrives, see Fig.~\ref{fig:sketch}(a). In this case
the effect of each pulse can be described by the classical PRC $Z(\varphi)$,
which determines the resulting phase shift given that the pulse arrived
at the phase $\varphi$. Another case when the PRCs are useful is
when the forcing is continuous in time but weak (Fig.~\ref{fig:sketch}(b)).
In this case the system remains close to the limit cycle, and the
phase dynamics can be described by the so-called infinitesimal phase
response curve \citep{Galan2005}. 

Therefore, the PRC-based approach is applicable for either weak or
sparse stimulation. However, in many realistic situations the stimuli
can be strong and frequent. In this case the system pushed away from
the limit cycle by one pulse does not return to it by the next pulse
arrival (Fig.~\ref{fig:sketch}(c)). In such situation, the usual
PRC can not account for the effect of the pulse, and a different approach
must be used. 

In this work we develop a framework for calculation of the oscillator
phase response to a series of pulses. The suggested approach is particularly
useful when the pulses are frequent or/and strong. In this case the
knowledge of the phase at which the pulse arrives does not allow to
calculate the phase shift it causes, so that the standard PRC is not
applicable. However, we show that the phase shift can still be calculated
using the phases at which\textit{ several} last pulses arrived. We
call the corresponding function ``phase response function'' (PRF).
We show that the impact of the previous pulses in the PRF falls exponentially
with time, which agrees with the experimental evidence that neurons
have exponentially decaying memory for past simulations \citep{Vardi2015,Goldental2015}. 

The necessity to overcome the limitations of the standard PRC have
been recognized previously. As a result, extensions for the PRC have
been proposed in \citep{Achuthan2009,Oprisan2004,Canavier1999,Guillamon2009,Castej=0000F3n2013,Wedgwood2013}.
In particular, in \citep{Achuthan2009,Oprisan2004,Canavier1999},
a phenomenological second order PRC was introduced that characterizes
the effect that the pulse has on the next cycle beyond the one containing
the perturbation. In \citep{Guillamon2009,Castej=0000F3n2013} the
authors introduced the ``amplitude response functions'' to capture
the system's response depending on the phase and the distance to the
cycle. The authors developed numerical algorithm to calculate phase-amplitude
response functions which constitutes an extension of the adjoint method
for PRCs \citep{Ermentrout1991}. Somewhat similar but distinct approach
was used in \citep{Wedgwood2013} where the authors used a transformation
to a moving orthonormal coordinate system around the limit cycle.

\begin{figure}[!t]
\begin{centering}
\includegraphics{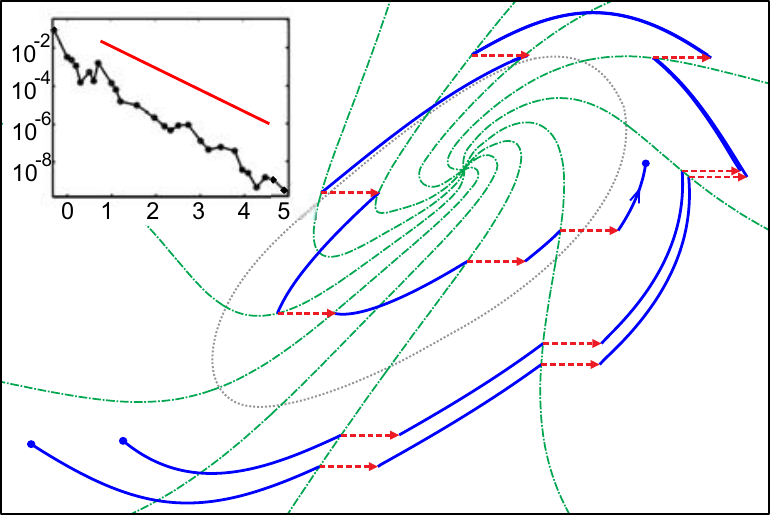}
\par\end{centering}
\caption{\label{fig:phaseplane}Dynamics of the FitzHugh-Nagumo oscillator
receiving pulses at given phases. Gray dotted line shows the limit
cycle, green solid dash-dotted lines the isochrons. Two trajectories
with different initial conditions are depicted on the phase plane
by blue lines, red dashed lines depict the action of the external
pulses. Inset: phase difference $\delta\varphi$ versus the phase
$\varphi_{k}$. Red line shows the slope $\delta\varphi\sim\mu^{\varphi}$.}
\end{figure}

In contrast to the previous works, our approach is not limited by
the number of significant pulses or the system dimension. The PRF
can be computed numerically or measured experimentally for oscillators
of arbitrary nature. 

The rest of the letter is organized as follows. First, we remind the
classical PRC model and introduce the concept of PRF. Then we show
how the PRF can be calculated and illustrate it for different oscillatory
systems. Finally, we report examples, where the PRF appropriately
models dynamics of forced or coupled systems whereas the classical
PRC fails.

To start with, we remind the classical PRC-based approximation of
an oscillatory system with pulse input. The oscillator is described
by the phase $\varphi$ which grows uniformly with $d\varphi/dt=\omega$
except for the time moments $t_{j}$ when the pulses arrive. At these
moments, the phase is shifted as
\begin{equation}
\varphi_{j}^{+}=\varphi_{j}^{-}+Z\left(\varphi_{j}^{-}\right),\label{eq:Z:def}
\end{equation}
where $Z(\varphi)$ is the PRC, and $\varphi_{j}^{-}$, $\varphi_{j}^{+}$
are the phases just before and after the pulse arrival at $t_{j}$.
The PRC-based approach provides a significant simplification comparing
to the study of large realistic systems, since the phase model is
one-dimensional, and the effects of the pulses are taken into account
discretely at points $t_{j}$.

The standard PRC approximation (\ref{eq:Z:def}) is valid in the case
of weak or sparse pulses. In this work we show that for strong or
frequent pulses, the phase shift caused by each pulse at $t_{j}$
can be approximated as 
\begin{equation}
\varphi_{j}^{+}=\varphi_{j}^{-}+Z_{n}\left(\varphi_{j-n+1}^{-},\dots,\varphi_{j}^{-}\right).\label{eq:Zn:def}
\end{equation}
Here the new function $Z_{n}:\mathbb{R}^{n}\mapsto\mathbb{R}$ is
the phase response function (PRF), $\varphi_{j}^{+}$ is the phase
just after the pulse at $t_{j},$ and $\varphi_{k}^{-}$ are the phases
just before the $k$-th pulse arrival at $t_{k}$. Hence, the phase
shift is determined by the phases at which the last $n$ pulses arrived.
Effectively, this means the emergence of the dynamical memory: the
impact of the current pulse becomes dependent on several previous
ones. The number $n$ of the significant pulses depends on how strong
and how frequent pulses are. In the case of weak or sparse pulses
$n=1$ and the PRF model (\ref{eq:Zn:def}) turns into the PRC model
(\ref{eq:Z:def}). 

Thus, the standard PRC is just a particular case of the PRF when the
stimulation is weak or sparse. Similarly with PRC, the PRF can be
measured numerically or experimentally for an arbitrary oscillator.
The direct method to obtain $Z_{n}(\varphi_{1},\varphi_{2},...,\varphi_{n})$
is to stimulate the oscillator by $n$ pulses at the phases $\varphi_{1}$,
...,$\varphi_{n}$ and measure the resulting phase shift. We emphasize
that the stimulation should be performed at the specified \emph{phases},
not times. Therefore, the evaluation of the phase after each stimulation
is necessary. The detailed description of a possible protocol is given
in the Supplemental Material. 

Thus, the PRF for a train of any number $n$ of pulses can be directly
obtained numerically or experimentally. However, the message of this
letter goes beyond this fact \textendash{} we show that only \textit{several}
recent pulses are significant. The qualitative explanation of this
feature is the following. The dynamics of every realistic oscillating
system can be split into the phase and the ``amplitude'' variables,
whereas the latter give the distance to the limit cycle. The phase
variable is neutrally stable (goldstone mode), while the amplitude
dynamics possesses contracting properties in average being in the
domain of attraction of the limit cycle. Therefore, the system ``forgets''
the amplitude variables after a sufficient period of time. In other
words, all orbits that are stimulated at the same phases approach
each other asymptotically, see blue orbits in Fig.~\ref{fig:phaseplane}.

This idea can be elaborated more precisely for the case of a 2-dimensional
system with a stable limit cycle. Following the approach developed
in \citep{Guillamon2009,Cabre2005}, such system can be reduced to
\begin{equation}
\dot{\varphi}=\omega,\qquad\dot{\rho}=\lambda\rho\label{eq:1}
\end{equation}
in a neighborhood of the cycle using a nonlinear coordinate transformation.
Here $\varphi\in\mathbb{R}(\mod1)$ is the phase of the system, $\omega=T^{-1}$
is the frequency ($T$ period), $\lambda<0$ is the Floquet exponent,
and the variable $\rho$ characterizes the distance to the limit cycle.

\begin{figure}
\begin{centering}
\includegraphics{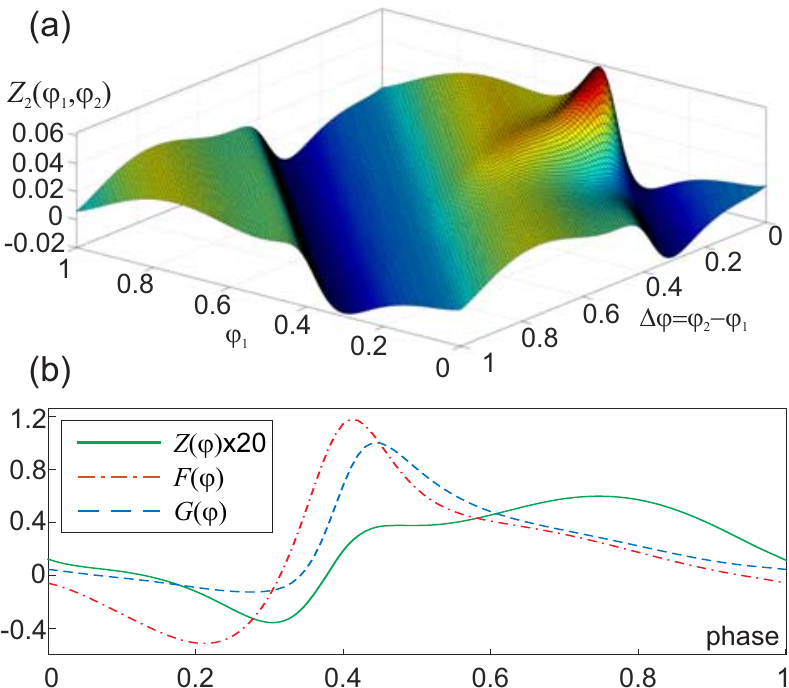}
\par\end{centering}
\caption{\label{fig:fu}(a) The PRF $Z_{2}(\varphi_{1},\varphi_{2})$ of the
Morris-Lecar model. (b) The standard PRC $Z(\varphi)$ (green solid
line), the function $F(\varphi)$ (red dash-dot line) and the function
$G(\varphi)$ (blue dashed line) of the Morris-Lecar model. }
\end{figure}

The effect of a short pulse on the oscillator (\ref{eq:1}) can be
given by a map $(\varphi,\rho)\to\left(\varphi^{*},\rho^{*}\right)$
which may be expressed in the form of power series 
\begin{eqnarray}
\varphi^{*} & = & \phi+\varepsilon P(\varphi)+\varepsilon^{2}Q(\varphi)+\varepsilon\rho F(\varphi)+\mathcal{O}(\varepsilon^{3}),\label{eq:phistar}\\
\rho^{*} & = & \rho+\varepsilon G(\varphi)+\mathcal{O}(\varepsilon^{2}),\label{eq:rhostar}
\end{eqnarray}
where $\varepsilon$ is the pulse strength and $P(\varphi)$, $Q(\varphi)$,
$F(\varphi)$, $G(\varphi)$ are period-1 functions. Here, we assume
that $\rho$ is of the order $\varepsilon$. Note that when the oscillator
is on the limit cycle ($\rho=0$), the phase shift caused by the pulse
equals to $Z(\varphi)=\varepsilon P(\varphi)+\varepsilon^{2}Q(\varphi)$,
which is the standard PRC. 

Now consider $n$ pulses arriving at phases $\varphi_{1}<\cdots<\varphi_{n}$
and determine the effect of this pulse train, namely the final phase
$\varphi_{n}^{*}$ after the last pulse arrival. If the pulses come
sparsely and the oscillator returns to the limit cycle by the arrival
of each pulse, the standard PRC may be used. In this case only the
last pulse matters, and $\varphi_{n}^{*}=\varphi_{n}+Z(\varphi_{n})$.
However, if the pulses are more frequent, the influence of the earlier
pulses is not negligible. 

To find the final phase in this case, consider the dynamics of the
oscillator during the whole pulse train. Each pulse causes the instant
shift according to (\ref{eq:phistar})\textendash (\ref{eq:rhostar}),
and between the pulses the system evolves according to (\ref{eq:1}).
This allows to construct a map that transforms the distance $\rho_{k}$
\textit{before} the $k$-th pulse into the distance $\rho_{k+1}$\textit{
before} the $(k+1)$-st pulse:
\begin{equation}
\rho_{k+1}=\left(\rho_{k}+\varepsilon G(\varphi_{k})\right)\mu^{\varphi_{k+1}-\varphi_{k}}+\mathcal{O}(\varepsilon^{2}),\label{eq:rhok}
\end{equation}
where $\mu=e^{\lambda T}$ is the multiplier of the limit cycle. Applying
(\ref{eq:rhok}) for $k=1,...,n-1$ and substituting the resulting
expression for $\rho_{n}$ into (\ref{eq:phistar}) gives
\begin{eqnarray}
\varphi_{n}^{*} & = & \varphi_{n}+\varepsilon P(\varphi_{n})+\varepsilon^{2}Q(\varphi_{n})+\varepsilon\rho_{1}F(\varphi_{n})\mu^{\varphi_{n}-\varphi_{1}}\nonumber \\
 & + & \varepsilon^{2}F(\varphi_{n})\sum_{k=1}^{n-1}G(\varphi_{k})\mu^{\varphi_{n}-\varphi_{k}}+O(\varepsilon^{3}).\label{eq:phinstar}
\end{eqnarray}

Note that the resulting phase depends on the initial distance $\rho_{1}$.
However, the term with $\rho_{1}$ decreases exponentially as the
interval between the first and the last pulses increases. The rate
of its decay is determined by the multiplier $\mu$, the term decays
significantly for $\varphi_{n}-\varphi_{1}\apprge\Theta$, where $\Theta=1/\left|\ln\mu\right|=1/\left|\lambda T\right|$.
In this case the influence of the initial distance $\rho_{1}$ is
negligible.

This result is illustrated in Fig. \ref{fig:phaseplane} where the
dynamics of the same oscillator with different initial conditions
is illustrated on the phase plane. If the same pulse trains are applied
at the same phases $\varphi_{k}$ then the trajectories converge after
a short transient.

The above allows us to say that if the pulse train is long enough
($\varphi_{n}-\varphi_{1}\apprge\Theta$), the final phase $\varphi_{n}^{*}$
depends only on the phases $\varphi_{k}$ of the incoming pulses and
does not depend on the prehistory $\rho_{1}$. In this case the PRF
is given by the phase shift $Z_{n}(\varphi_{1},...,\varphi_{n})=\varphi_{n}^{*}-\varphi_{n}$
that can be approximated as 

\begin{equation}
Z_{n}(\varphi_{1},...,\varphi_{n})=Z(\varphi_{n})+\varepsilon^{2}F(\varphi_{n})\sum_{k=1}^{n-1}G(\varphi_{k})\mu^{\varphi_{n}-\varphi_{k}}.\label{eq:Zn}
\end{equation}
Note that for strong attraction $\mu\to0$ the PRF transforms to the
standard PRC $Z(\varphi_{n}).$ For the finite attraction $\mu>0$,
the effect of the past pulses decays exponentially, therefore only
those pulses matter whose phases fall into the interval $\varphi_{n}-\varphi_{k}\apprle\Theta$.
Hence, the number of the pulses to be taken into account can be estimated
as $n\sim f/\left|\lambda T\right|$, where $f$ is the typical frequency
at which pulses arrive.

The above analysis not only allowed to estimate the number of significant
pulses in the train, but also provides an approximate formula (\ref{eq:Zn})
for the PRF. The expression (\ref{eq:Zn}) suggests that the system
response may be divided into two contributions. The first one is the
impact of the current pulse captured by the standard PRC $Z(\varphi_{n})$.
The second one represents the ``correction'' to the PRC due to the
impact of the previous pulses. To verify the developed theory we stimulated
various oscillators by doublets of pulses at different phases $\varphi_{1}$
and $\varphi_{2}$ and measured the PRF $Z_{2}(\varphi_{1},\varphi_{2})$
directly. Then we checked whether the correction term is given by
$Z_{2}(\varphi_{1},\varphi_{2})-Z(\varphi_{1})=\varepsilon^{2}F(\varphi_{2})G(\varphi_{1})\mu^{\varphi_{2}-\varphi_{1}}$
as it should be according to (\ref{eq:Zn}). Our tests for several
popular oscillatory models \textendash{} FitzHugh-Nagumo, Morris-Lecar,
Hodgkin-Huxley and Van-der-Pol showed remarkable accuracy of this
approximation. The details of the protocol and the models are given
in the Supplemental Material. 

The results of the simulations allowed us to construct the functions
$F(\varphi)$ and $G(\varphi)$ for the tested oscillators, see Fig.~\ref{fig:fu}
and Figs. S1-S4. It is remarkable that the function of $n$ variables
$Z_{n}(\varphi_{1},...,\varphi_{n})$ can be approximated by functions
of a single variable. Moreover, although the approximation (\ref{eq:Zn})
is derived for 2D oscillators, it can be practically applicable for
higher-dimensional systems, as the example of the Hodgkin-Huxley model
shows. A presumable reason for that is the existence of the so-called
leading manifold of a stable limit cycle \citep{Shilnikov2001} on
which the dynamics is governed by (\ref{eq:1}). 

\begin{figure}
\centering{}\includegraphics{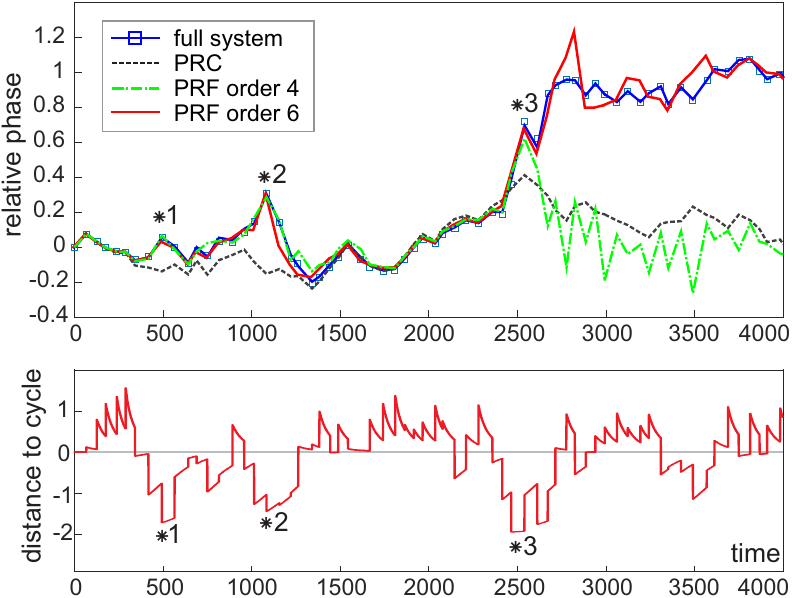}\caption{\label{fig:vdp}The dynamics of the forced Van-der-Pol oscillator
for $\alpha=0.01$, $\varepsilon=1$. The top panel shows the relative
phase $\varphi'=\varphi-\omega t$, the bottom panel the distance
from the limit cycle (positive outside of the cycle and negative inside).
On the top panel, the blue solid line with bars shows the dynamics
of the full model, black dashed line the PRC approximation, green
dash-dot line the approximation by the PRF of the fourth order, red
solid line the PRF of the sixth order. }
\end{figure}

In the following we show examples where the PRF shows essential advantages
comparing to PRC. Consider the Van-der-Pol oscillator 
\begin{equation}
\ddot{x}-\alpha(1-x^{2})\dot{x}+x=0,\label{eq:VdP}
\end{equation}
stimulated by pulses that instantly change the variable $x$ to $x+\varepsilon$.
For small $\alpha$, the phase $\varphi$ can be introduced geometrically,
see \citep{Guckenheimer2013,Nekorkin2015} and Supplemental Material.
Figure~\ref{fig:vdp} illustrates the dynamics of (\ref{eq:VdP})
under the action of a pulse train. The pulses are applied at random
moments with the inter-pulse intervals distributed homogeneously within
the limits $[40,80]$. 

One can observe that the PRF approach provides accurate results for
the phase dynamics even in the case when the PRC fails. In particular,
we compare the results obtained by using the PRF of the sixth order
(taking into account 6 last pulses), the fourth order, and the standard
PRC. One may see that the standard PRC is sometimes effective, but
at certain moments it becomes inaccurate. Particularly, it gives substantial
errors at $t\approx500$ and $t\approx1100$ (asterisks 1 and 2 on
Fig. \ref{fig:vdp}), and becomes absolutely inapplicable at $t\approx2500$
(asterisk 3) when the error exceeds one. The bottom panel reveals
that the PRC fails in the moments when the oscillator goes far from
the limit cycle. The same happens with the PRF of the 4-th order.
In contrast, the PRF of the 6-th order provides correct results.

\begin{figure}
\centering{}\includegraphics{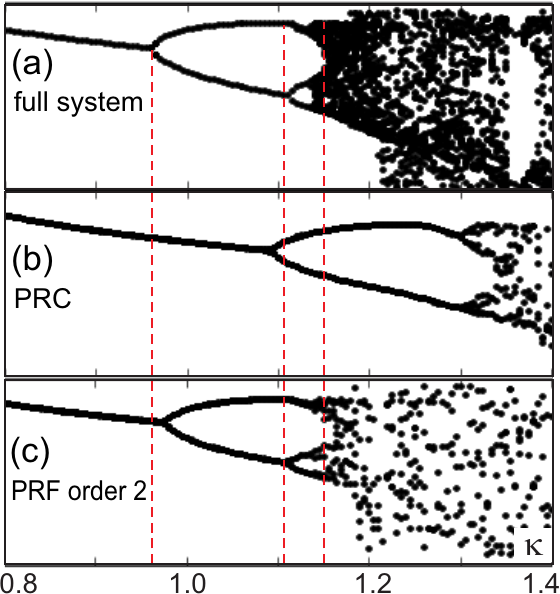}\caption{Bifurcation diagram for two coupled Van-der-Pol oscillators and approximations
using PRF, $\alpha=2$. (a) Original numeric bifurcation diagram.
(b) Diagrams obtained using the approximations with the PRC. (c) Diagram
obtained with the PRF of order 2. \label{fig:BD-VDP}}
\end{figure}

As a final demonstration of the PRF advantage, we present a bifurcation
diagram in Fig.~\ref{fig:BD-VDP} for the two Van-der-Pol oscillators
(\ref{eq:VdP}) with pulse coupling. The coupling is organized as
follows. When the first oscillator crosses the threshold $x_{1}=0$
from below, the pulse is sent to the second oscillator. The latter
is then instantly perturbed so that $x_{2}^{+}=x_{2}^{-}+\kappa x_{2}^{-}$.
Similarly, when the second oscillator crosses the threshold, the pulse
is sent to the first one.The PRC approximation (Fig.~\ref{fig:BD-VDP}(b))
provides qualitatively correct results showing the transition to chaos
through a period-doubling cascade. However, the bifurcation points
differ significantly from the full system. In contrast, the PRF approach
allows to predict these transitions much more accurately (Fig.~\ref{fig:BD-VDP}(c)).

To conclude, the concept of PRF provides a novel approach for the
modeling of forced or coupled oscillators. It preserves the advantages
of the standard PRC-based approach: low dimensionality of the model,
computational effectiveness, and the possibility to obtain the PRF
for an arbitrary oscillator. At the same time, the PRF remains valid
for stronger and more frequent stimulation and thus may capture essentially
new dynamical effects. The latter point allows to consider the PRF
not only as approximation of full models, but also as a stand-alone
model and a test-bed for new phenomena and hypotheses.

Natural application of the PRF is simulation of coupled oscillators
\citep{Mirollo1990,Bottani1995,Herz1995,Vreeswijk1996,Bressloff1997,Izhikevich1999,Goel2002,Neves2009},
such as large populations of neurons \citep{Abbot1993,Coombes1997,Zillmer2007,Olmi2010,Taylor2011},
ensembles of chemical \citep{Bar2012,Horvath2012,Vanag2016,Chen2016},
electronic \citep{Klinshov2014,Klinshov2015} or optic \citep{Vladimirov2005}
oscillators. The PRF may be generalized for pulses of different amplitude,
which allows to account for inhomogeneous distribution of synaptic
weights \citep{Teramae2012,Klinshov2014Fukai}. Coupling delays can
also be easily included \citep{Ernst1995,Gerstner1996,Timme2002a,Zeitler2009,Klinshov2013}.
Thus, the PRF provides an effective tool for simulation of oscillatory
networks which we hope will be demanded in neuroscience and other
fields.
\begin{acknowledgments}
The research was supported by the Russian Scientific Foundation (project
16-42-01043 for the Institute of Applied Physics) and the German Research
Foundation (project SCHO 307/15-1 and YA 225/3-1 for the TU Berlin).
We also acknowledge the valuable discussions with Dr. Leonhard L\"ucken.
\end{acknowledgments}

\part*{Supplementary materials}

\renewcommand{\theequation}{S\arabic{equation}}

\renewcommand{\thefigure}{S\arabic{figure}}

\section{The protocol for direct measurement of the PRF}

Here we demonstrate how the PRF can be calculated numerically or measured
experimentally. Assume we have an oscillator with a stable limit cycle
$\gamma$ and we need to obtain the PRF of the order $n$ at a certain
point $Z_{n}(\varphi_{1},...,\varphi_{n})$. First we define a specific
point $O$ on the cycle with the phase $\varphi_{0}=0$. Then the
phase on the limit cycle is well defined as 

\begin{equation}
\varphi=\omega(t-t_{0}),
\end{equation}
where $t$ is a current time, $t_{0}$ is the moment of the last passage
of point $O$, and $\omega=T^{-1}$, where $T$ is the period of the
limit cycle.

On the first step, we set the oscillator to point $O$ at $t=0$,
or just wait until it reaches this point. The we apply one pulse at
the moment $t_{1}=T\varphi_{1}$. The phase of the oscillator at the
pulse arrival equals $\varphi_{1}$. Then we skip a long enough transient
assuring the convergence to the limit cycle and measure the phase
$\varphi_{c}$ at some moment $t_{c}$. The phase of the unperturbed
system would be $\omega t_{c}(\mbox{\,mod\,}1)$, thus we can calculate
the phase shift which equals the PRF of the first order

\begin{equation}
Z_{1}(\varphi_{1})=\varphi_{c}-\omega t_{c}\,\,(\mbox{\,mod\,}1).
\end{equation}

On the second step, we reset the oscillator again to point $O$ at
$t=0$ and apply two pulses, the first one at the moment $t_{1}=T\varphi_{1}$,
and the second one at the moment $t_{2}=T(\varphi_{2}-Z_{1}(\varphi_{1}))$.
The phase of the oscillator equals $\varphi_{1}$ at the first pulse
arrival, and $\varphi_{2}$ at the second pulse arrival. Thus, measuring
the phase $\varphi_{c}$ at moment $t_{c}$ after a long transient,
we obtain the PRF of the second order

\begin{equation}
Z_{2}(\varphi_{1},\varphi_{2})=\varphi_{c}-\omega t_{c}(\mod1).
\end{equation}

Continuing this iterative process one may measure the PRF of any order.

\newpage{}

\section{The impact and the sensitivity functions of various oscillators}

To measure the impact and the sensitivity functions of an oscillator,
we calculated the second order PRF $Z_{2}(\varphi_{1},\varphi_{2})$
on a grid $(\varphi_{1},\varphi_{2})$. Subtracting the standard PRC
allows to find the ``correction'' $\Delta Z(\varphi_{1},\varphi_{2})=Z_{2}(\varphi_{1},\varphi_{2})-Z(\varphi_{1})$
which should equal according to Eq. (7) of the main text
\begin{equation}
\Delta Z(\varphi_{1},\varphi_{2})=Z_{2}(\varphi_{1},\varphi_{2})-Z(\varphi_{1})=\varepsilon^{2}F(\varphi_{2})G(\varphi_{1})\mu^{\varphi_{2}-\varphi_{1}}.
\end{equation}
A simple and readily verified consequence is that $\Delta Z(\varphi_{1},\varphi_{2}+1)=\mu\Delta Z(\varphi_{1},\varphi_{2})$.
We checked that this equality indeed holds with high accuracy, which
allows to estimate the value of $\mu$. After that the functions $F(\varphi)$
and $G(\varphi)$ can be estimated as follows. On the first step,
we fix $\varphi_{1}$, change $\varphi_{2}$ from zero to one and
calculate $F(\varphi_{2})=k\Delta Z(\varphi_{1},\varphi_{2})\mu^{-\varphi_{2}}$.
Here $k$ is a normalization coefficient selected so that the maximal
absolute value of $F(\varphi)$ equals one. On the second step, we
fix $\varphi_{2}$ , change $\varphi_{1}$ from zero to one and use
the previously calculated values of $F(\varphi_{2})$ to calculate
$G(\varphi_{1})=\varepsilon^{-2}\Delta Z(\varphi_{1},\varphi_{2})\mu^{\varphi_{1}-\varphi_{2}}/F(\varphi_{2})$. 

The suggested algorithm has been applied to several oscillatory systems
- Moris-Lecar, FitzHugh-Nagumo, Van-der-Pol and Hodgkin-Huxley. Below
the details of the models are given and the results are shown. By
different colors are plotted the functions $F(\varphi_{2})$ obtained
for different $\varphi_{1}$ and the functions $G(\varphi_{1})$ obtained
for different $\varphi_{2}$. Note that the difference is small.

\begin{figure}[H]
\begin{centering}
\includegraphics{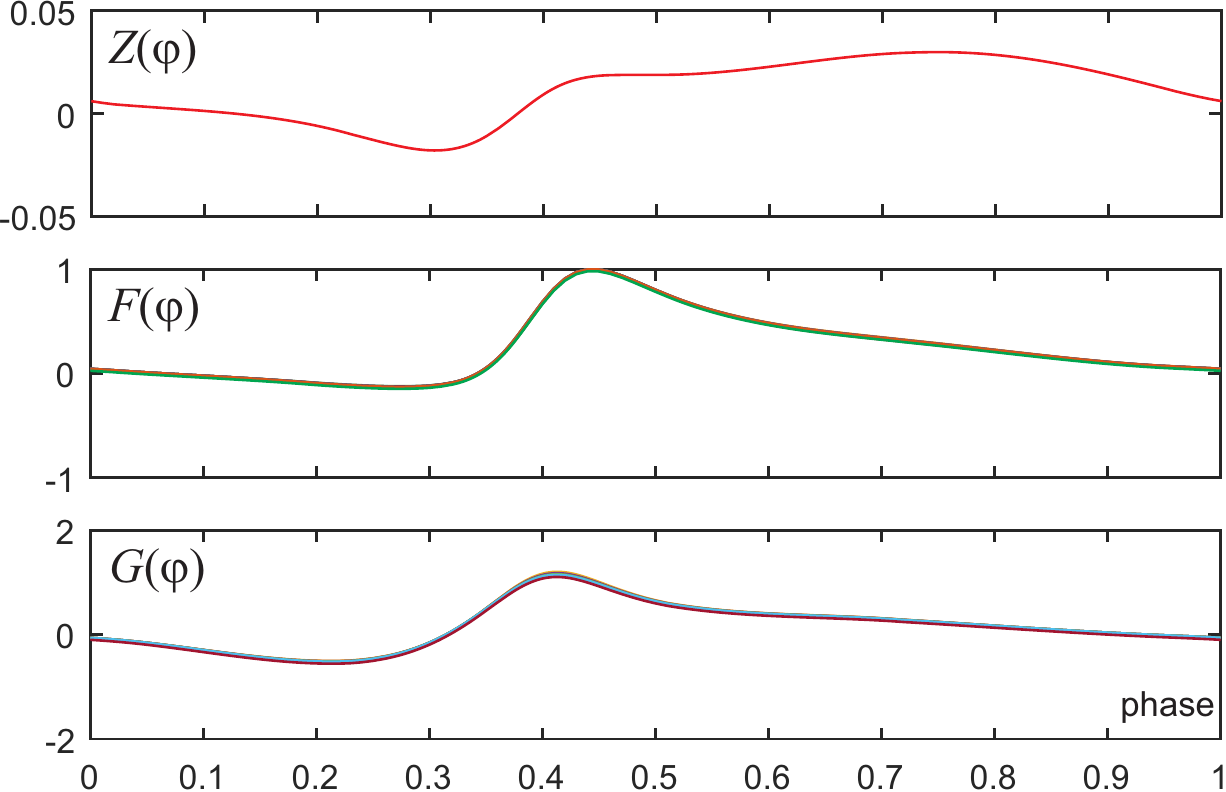}
\par\end{centering}
\caption{The results for the Morris-Lecar model.}
\end{figure}

The Morris-Lecar model \citep{Morris1981} is given by the system

\begin{eqnarray}
C_{M}\frac{dV}{dt} & = & I-g_{L}\left(V-V_{L}\right)-g_{K}m\left(V-V_{K}\right)-g_{Ca}\frac{1+\tanh\frac{V-V_{1}}{V_{2}}}{2}\left(V-V_{Ca}\right),\\
\frac{dm}{dt} & = & \phi\cosh\frac{V-V_{3}}{2V_{4}}\left(\frac{1+\tanh\frac{V-V_{3}}{V_{4}}}{2}-m\right).
\end{eqnarray}

with $I=100$, $C_{M}=50$, $g_{Ca}=4$, $g_{K}=8$, $g_{L}=2$, $V_{Ca}=120$,
$V_{K}=-80$, $V_{L}=-60$, $\phi=0.04$, $V_{1}=-1.2$, $V_{2}=18$,
$V_{3}=10$, $V_{4}=17$. The pulse instantly changes $V$ by the
value $\Delta V=2$. The results are given in Fig. 3 of the main text
and Fig. S1.

\begin{figure}[H]
\begin{centering}
\includegraphics{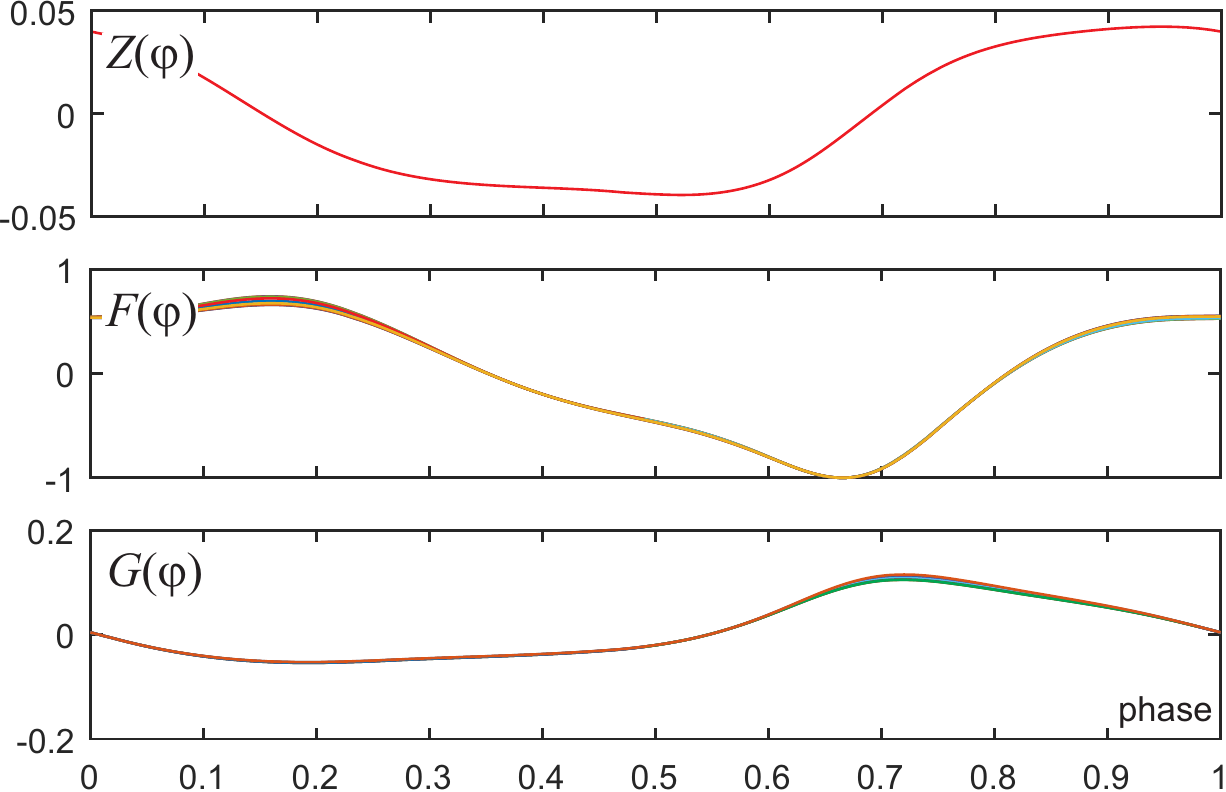}
\par\end{centering}
\caption{The results for the FitzHugh-Nagumo model.}
\end{figure}

The FitzHugh-Nagumo model \citep{FitzHugh1961,Nagumo1962} is given
by the system

\begin{eqnarray}
\frac{dv}{dt} & = & I+v-\frac{v^{3}}{3}-u,\\
\frac{du}{dt} & = & a\left(v+b-cu\right).
\end{eqnarray}

with $I=1$, $a=0.8$, $b=0.7$, $c=0.8$. The pulse instantly changes
$v$ by the value $\Delta v=0.2$. The results are given in Fig. S2.

\begin{figure}[H]
\begin{centering}
\includegraphics{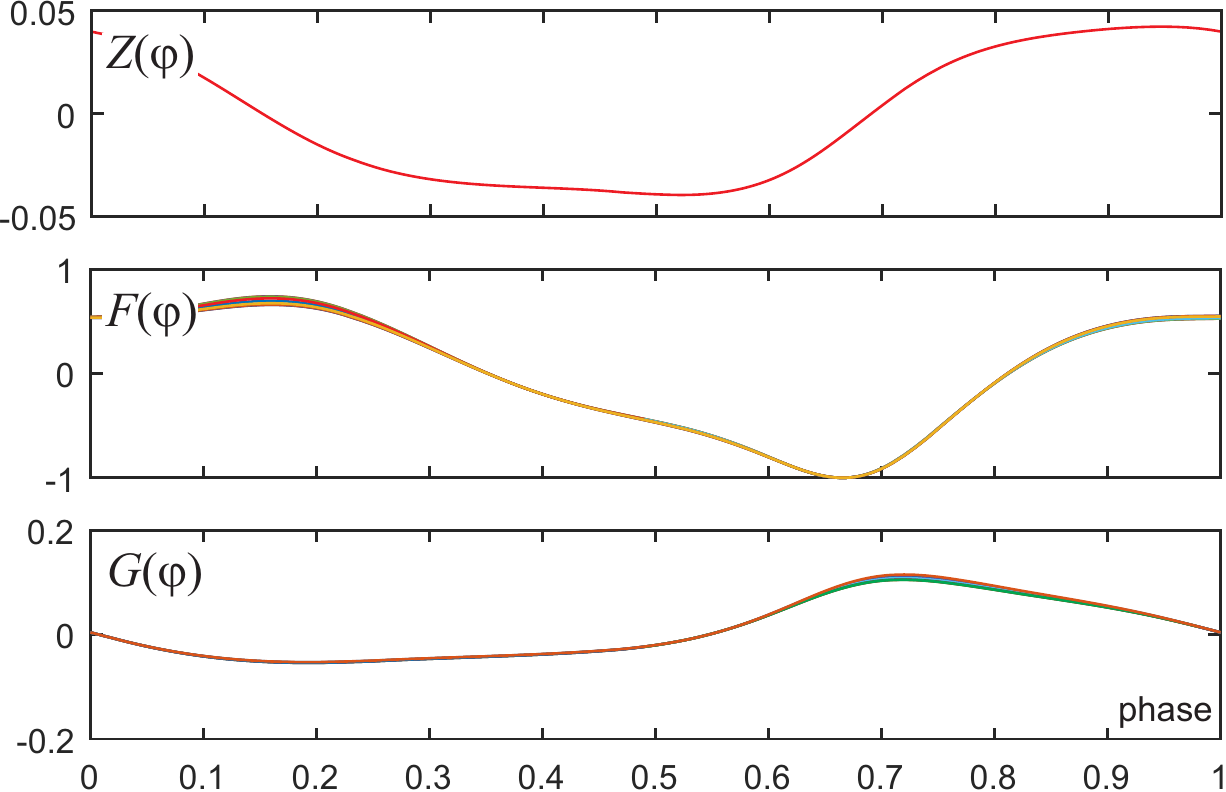}
\par\end{centering}
\caption{The results for the Van der Pol model.}
\end{figure}

The Van-der-Pol oscillator \citep{VdP1926} is given by the equation 

\begin{equation}
\frac{d^{2}x}{dx^{2}}-\alpha\left(1-x^{2}\right)\frac{dx}{dt}+x=0
\end{equation}

with $\alpha=0.2$. The pulse changes $x$ by the value $\Delta x=0.5$.
The results are given in Fig. S3.

\newpage{}

\begin{figure}[H]
\begin{centering}
\includegraphics{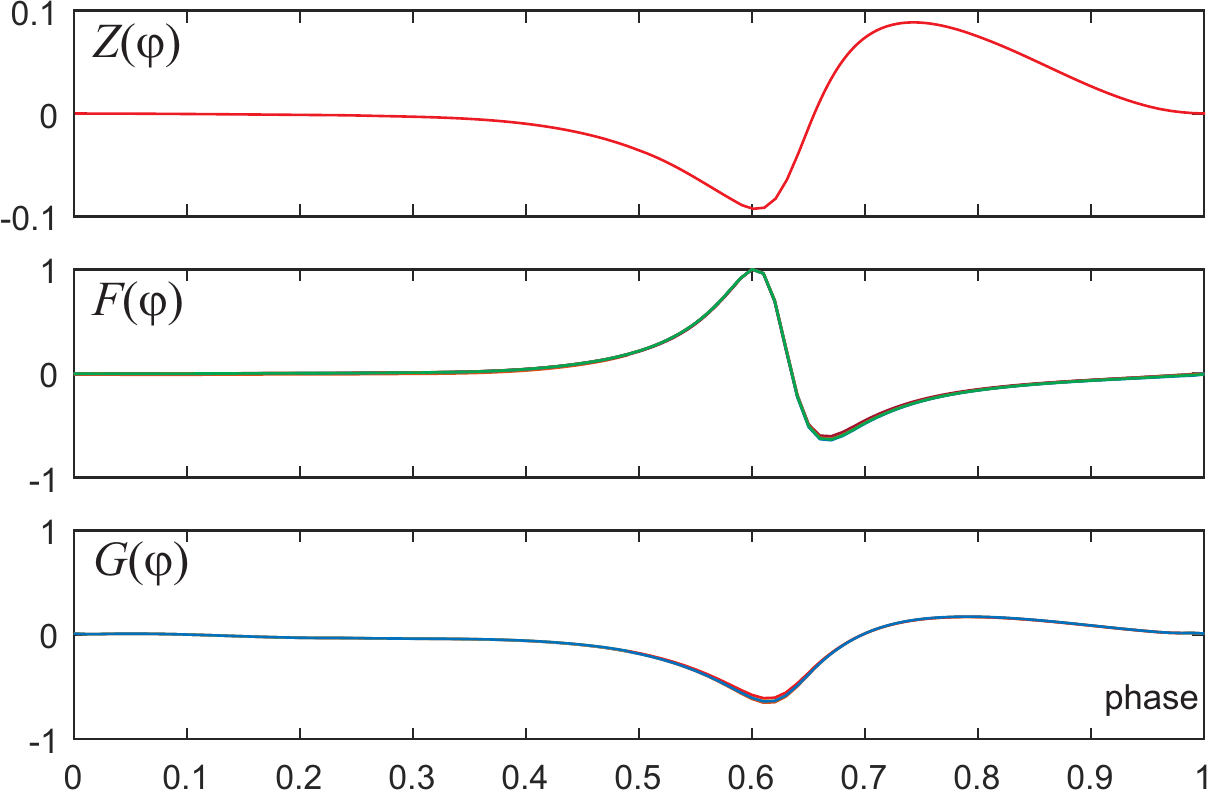}
\par\end{centering}
\caption{The results for the Hodgkin-Huxley model.}
\end{figure}

The Hodgkin-Huxley model \citep{HH1952} is given by the system

\begin{eqnarray}
C_{M}\frac{dV}{dt} & = & I-g_{K}n^{4}\left(V-V_{K}\right)-g_{Na}m^{3}h\left(V-V_{Na}\right)-g_{l}\left(V-V_{L}\right),\\
\frac{dn}{dt} & = & \frac{0.01V+0.55}{1-\exp\left(-0.1V-5.5\right)}\left(1-n\right)-0.125\exp\left(\frac{-V-65}{80}\right)n,\\
\frac{dm}{dt} & = & \frac{0.1V+4}{1-\exp\left(-0.1V-4\right)}\left(1-m\right)-4\exp\left(\frac{-V-65}{18}\right)m,\\
\frac{dh}{dt} & = & 0.07\exp\left(\frac{-V-65}{20}\right)\left(1-h\right)-\frac{1}{1+\exp\left(-0.1V-3.5\right)}h
\end{eqnarray}

with $C_{M}=1$, $I=10$, $g_{Na}=120$, $g_{K}=36$, $g_{L}=0.3$,
$V_{Na}=50$, $V_{K}=-77$, $V_{l}=-54.5$. The pulse changes $V$
by $\Delta V=3$. The results are given in Fig. S4.

\newpage{}

\section{The analysis of the Van-der-Pol oscillator}

First, we rewrite (9) of the main text as 

\begin{eqnarray}
\frac{dx}{dt} & = & y,\\
\frac{dy}{dt} & = & \alpha y(1-x^{2})-x.
\end{eqnarray}
Then, we introduce the polar coordinates $\varphi$ and $R$ so that
$x=R\cos2\pi\varphi$ and $y=-R\sin2\pi\varphi$. Substituting this
into (2),(3) results in

\begin{eqnarray}
\dot{R}\cos2\pi\varphi-2\pi\dot{\phi}R\sin2\pi\varphi & = & -R\sin2\pi\varphi,\\
-\dot{R}\sin2\pi\varphi-2\pi\dot{\varphi}R\cos2\pi\varphi & = & -\alpha R\sin2\pi\varphi\left(1-R^{2}\cos^{2}2\pi\varphi\right)-R\cos2\pi\varphi,
\end{eqnarray}
where the dot means is the derivative over time. From (4), (5) one
can express $\dot{R}$ and $\dot{\varphi}$ as 

\begin{eqnarray}
\dot{R} & = & \alpha R\sin^{2}2\pi\phi\left(1-R^{2}\cos2\pi\varphi\right),\\
\dot{\phi} & = & \frac{1}{2\pi}\left(1+\alpha\sin2\pi\varphi\cos2\pi\varphi\left(1-R^{2}\cos^{2}2\pi\varphi\right)\right).
\end{eqnarray}
Note that $\dot{\phi}\approx1$ for $\alpha\ll1$, and the right parts
of (6), (7) may be averaged \citep{Guckenheimer2013,Nekorkin2015}
leading to

\begin{eqnarray}
\dot{R} & = & \frac{1}{8}\alpha R\left(4-R^{2}\right),\\
\dot{\varphi} & = & \frac{1}{2\pi}.
\end{eqnarray}

Easy to see that the variable change $\rho=1-4/R^{2}$ transforms
(8), (9) into

\begin{eqnarray}
\dot{\varphi} & = & \omega,\\
\dot{\rho} & = & \lambda\rho
\end{eqnarray}
with $\omega=1/(2\pi)$, $\lambda=-\alpha$. Thus, for $\alpha\ll1$
the variable $\varphi$ is the oscillator's phase.

\newpage{}

\end{document}